\documentstyle[12pt,amssymb]{article}

\newcommand{\mrm}[1]{\mathrm{#1}}
\newmathalphabet*{\mbf}{cmr}{b}{n}
\newmathalphabet*{\mii}{cmr}{m}{it}
\newmathalphabet*{\mtt}{cmtt}{m}{n}

\newcommand{\mt}{m_{\mathrm{t}}}
\newcommand{\Gt}{\Gamma_{\mathrm{t}}}

\newcommand{\GW}{\Gamma_{\mathrm{W}}}

\newcommand{\alphas}{\alpha_{\mrm{s}}}

\newcommand{\pT}{p_{\perp}}

\renewcommand{\b}{\mathrm{b}}

\renewcommand{\d}{\mathrm{d}}
\newcommand{\e}{\mathrm{e}}

\newcommand{\g}{\mathrm{g}}

\newcommand{\p}{\mathrm{p}}
\newcommand{\q}{\mathrm{q}}

\renewcommand{\t}{\mathrm{t}}
\renewcommand{\u}{\mathrm{u}}

\newcommand{\B}{\mathrm{B}}

\newcommand{\K}{\mathrm{K}}

\newcommand{\T}{\mathrm{T}}
\newcommand{\W}{\mathrm{W}}
\newcommand{\Z}{\mathrm{Z}}
\newcommand{\bbar}{\overline{\mathrm{b}}}

\newcommand{\pbar}{\overline{\mathrm{p}}}
\newcommand{\qbar}{\overline{\mathrm{q}}}

\newcommand{\tbar}{\overline{\mathrm{t}}}

\newcommand{\Bbar}{\overline{\mathrm{B}}}

\newcommand{\ee}{\e^+\e^-}

\newcommand{\gammaZ}{\gamma^* / \Z^0}

\newcommand{\mmax}{\mrm{max}}

\newenvironment{Itemize}{\begin{list}{$\bullet$}%
{\setlength{\topsep}{0.2mm}\setlength{\partopsep}{0.2mm}%
\setlength{\itemsep}{0.2mm}\setlength{\parsep}{0.2mm}}}%
{\end{list}}
\newcounter{enumct}
\newenvironment{Enumerate}{\begin{list}{\arabic{enumct}.}%
{\usecounter{enumct}\setlength{\topsep}{0.2mm}%
\setlength{\partopsep}{0.2mm}\setlength{\itemsep}{0.2mm}%
\setlength{\parsep}{0.2mm}}}{\end{list}}
\newlength{\captivewidth}
\begin{document}

%set sloppy attitude to line breaks
\sloppy

\pagestyle{empty}

\begin{flushright}
CERN-TH.7199/94 \\
DTP/94/18
\end{flushright}

\vspace{\fill}

\begin{center}
{\LARGE\bf Colour Correlations and Multiplicities}\\[3mm]
{\LARGE\bf in Top Events}\\[10mm]
{\Large Valery A. Khoze} \\[3mm]
{\large Department of Physics, University of Durham} \\[1mm]
{\large Durham DH1 3LE, England} \\[5mm]
{\large and} \\[5mm]
{\Large Torbj\"orn Sj\"ostrand} \\[3mm]
{\large Theory Division, CERN} \\[1mm]
{\large CH-1211 Geneva 23, Switzerland} \\
\end{center}

\vspace{\fill}

\begin{center}
\bf{Abstract}
\end{center}
\vspace{-0.5\baselineskip}
\noindent
In events of the type $\ee \to \t\tbar \to \b\W^+ \, \bbar\W^-$,
particle production could depend in a non-trivial way on the kinematics
of the process. Energetic perturbative gluon radiation can be generated
(when kinematically allowed) by the original $\t\tbar$ system and by
the $\t \to \b\W^+$ and $\tbar \to \bbar\W^-$ decays, with negligible
interference between the production and decay stages and between
the $\t$ and $\tbar$ decays. Soft perturbative gluon emission and
non-perturbative fragmentation does introduce a correlation, however.
To illustrate the size of these effects, we study the multiplicity
as a function of the angle between the $\b$ and $\bbar$ jets, close
to the $\t\tbar$ threshold. Also potential uncertainties in top mass
determinations are briefly addressed.

\vspace{\fill}

\noindent
CERN-TH.7199/94 \\
March 1994

\clearpage

\pagestyle{plain}
\setcounter{page}{1}

\section{Introduction}

In high-energy physics, ``tomorrow belongs'' to the detailed study of
heavy unstable particles ($\Z$ and $\W$ bosons, top quarks,
SUSY particles, \ldots). An important aim of future experiments is
the precise determination of their parameters, primarily their
masses. This requires a detailed understanding of production and
decay mechanisms (including interference effects) and, in
particular, of the effects arising from
the large width of many of these objects,
$\Gamma \sim {\cal O}(1 \; \mrm{GeV})$.
One can find a long list
of examples in Refs. \cite{K1,K2,K3}. For instance, in Ref. \cite{K2}
the present authors have examined in detail the QCD rearrangement
phenomena in hadronic $\W^+\W^-$ events and their impact on the $\W$
mass reconstruction, both for perturbative and non-perturbative
effects. Here we concentrate on another topical problem, namely the
QCD interconnection phenomena that may occur when a top quark and
top antiquark decay and hadronize close to each other in space and
time. The word `interconnection' is here introduced to cover those
aspects of final-state particle production that are not dictated
by the separate $\t$ and $\tbar$ decays, but can only be understood
in terms of a joint action of the two.

The indirect evidence for the existence of the top quark is very
strong. If it behaves as predicted by the minimal standard model,
then its mass $\mt$ is in the range of 120--200~GeV \cite{K7}.
Such a $\t$ quark has a good chance to be detected at the
Tevatron in the foreseeable future. The detailed study of the top
will be one of the most important physics topics at future linear
$\ee$ colliders \cite{K4,K6}.

The dominance of the $\t \to \b \W^+$ decay mode leads to a large
top width $\Gt$, which is about 1 GeV for a canonical mass
$\mt \simeq 160$~GeV. Since the top is heavier than 120~GeV, the width
$\Gt$ is larger than the typical hadronic scale $\mu \sim 1$~fm$^{-1}$,
and the top may decay before it has time to hadronize \cite{K9,K10}.
It is precisely the large width that makes top physics so unique,
see e.g.
\cite{K6,K10}. Firstly, the top decay width $\Gt$ provides an infrared
cut-off for the strong forces between the $\t$ and the $\tbar$
\cite{K11,K12}. The width $\Gt$ acts as a physical `smearing'
\cite{K13}, and the top production becomes a quantitative prediction
of perturbative QCD, largely independent of non-perturbative
phenomenological algorithms. Secondly, $\Gt$ controls the QCD
interferences between radiation occurring at different stages of the
$\t\tbar$ production processes \cite{K1,K14,K15}. These interferences
affect the structure of the colour flows
in the $\t\tbar$ events and may provide a potentially serious source of
uncertainties in the reconstruction of the final state.

The perturbative aspects of radiative interference phenomena are
complex but, in principle, well controllable. We speak of virtual
interference when
gluons are produced at one stage of the process and absorbed at another.
Real interference occurs as well, since the same real gluon can be
emitted from the different stages of the process.
Analogously to the $\W^+\W^-$ case \cite{K2} we
expect that the perturbative restructuring is suppressed because of
the space--time separation between the decays of the $\t$ and $\tbar$
quarks. However, the non-perturbative aspects are not so well
understood, and a priori there is no obvious reason why
interconnection effects
have to be small in the fragmentation process. Moreover, the
$\b$ and $\bbar$ coming from the top decays carry compensating colour
charges and therefore have to `cross-talk' in order to produce a
final state made up of colourless hadrons.

As a specific topical example, which will be used as a basis for the
continued discussion, we consider the production and decay of a
$\t\tbar$ pair in the process
\begin{equation}
\e^+\e^- \to \t\tbar \to \b\W^+ \, \bbar\W^-    ~.
\end{equation}
For simplicity we assume that the $\W$'s decay leptonically,
so the colour flow is generated only by the $\t$ and $\b$ quarks.
Further, we restrict ourselves to the region a few GeV above the
$\t\tbar$ threshold to exemplify the size of effects.
The radiation pattern is especially simple in
this region. At higher energies one has to take into account
the energetic gluon radiation from the $\t\tbar$ dipole (see
Ref. \cite{K14} and section 2 below). At the other extreme, just
near threshold, the radiation amplitude is strongly modified
by the QCD Coulomb-like interaction between the $\t$ and $\tbar$
\cite{K16,K17}.

The topology of the final state in process (1) was discussed in
Ref. \cite{K18} for a mass region $\mt \simeq 120$~GeV, where the
top fragmentation and decay times are comparable, and therefore the
non-perturbative dynamics has an especially rich
structure. With the current larger estimate for the top mass, and
in a region close to threshold, the dominance of the `1-string
topology' of Ref. \cite{K18} is total. The average energy lost to the
$\t\tbar$ string is $\Delta E \approx 2 \kappa v_{\t}
\gamma_{\t} \hbar / \Gt \approx 0.1$~GeV for $\mt = 160$~GeV and
$E_{\mrm{cm}} = 330$~GeV, with string tension
$\kappa \approx 1$~GeV/fm \cite{TS1}.
This means that no particles at all have time to form before the
$\t$ and $\tbar$ decay; the original $\t\tbar$ string is remembered
only as some small wrinkles on the $\b\bbar$ string formed after the
decays.

QCD interconnection effects undermine the traditional description of the
final state in process (1). It becomes impossible (even assuming the
complete reconstruction of the $\W$'s) to subdivide the hadronic final
state into two groups of particles, one of which is associated with the
top decay and the other with the antitop decay: some hadrons originate
from the joint action of the two systems.

The interplay of several particle production sources is reminiscent
of the colour rearrangement effects we have studied for
$\ee \to \W^+ \W^- \to \q_1\qbar_2 \, \q_3\qbar_4$ \cite{K2},
but there are important differences. From the onset, $\W^+ \W^-$
events consist of two separate colour singlets, $\q_1\qbar_2$ and
$\q_3\qbar_4$, so that there is no logical imperative of an interplay
between the two. Something extra has to happen to induce a colour
rearrangement to $\q_1\qbar_4$ and $\q_3\qbar_2$ singlets,
such as a perturbative exchange of gluons or a non-perturbative string
overlap. This introduces a sizeable dependence on the space--time picture,
i.e. on how far separated the $\W^+$ and $\W^-$ decay vertices are.
Except in the unlikely case that top-flavoured hadrons would have time
to form, the process $\ee \to \t\tbar \to \b\W^+\bbar\W^-$ only involves
one colour singlet. Therefore an interplay is here inevitable, while
a colour rearrangement of the above kind is impossible.

Analogously to the celebrated string/drag effect \cite{K19,K20},
the colour topology could induce azimuthal anisotropies
in the distribution of the particle flows (see also Ref. \cite{K18}).
In particular, the $\b$ jets from the $\t$ decays could become
azimuthally asymmetric. Such an asymmetry should depend on the
relative orientation of the $\b$ and $\bbar$ jets.

One of the main objectives of a future linear $\ee$ collider will
be to determine the top mass, which can be done in several different
ways. One method is to reconstruct the top invariant mass event by
event \cite{K4}, another is to measure the top momentum
distribution \cite{K3,K21}. In either case, the
colour flow restructuring discussed above could introduce
the potentiality for a systematic bias in the top mass determination.

It is not our intention to go through all the details of the problem.
In this paper we shall concentrate on the possible manifestations
of the QCD interconnection effects in the distribution of the particle
flow in the final state of process (1).
As a specific example, we study the total multiplicity of
double leptonic top decays as a function of the relative angle between
the $\b$ and $\bbar$ jets.

\section{The Perturbative Picture}

Within the perturbative scenario, the particle distribution in the
final state of an $\ee \to \t\tbar$ event is controlled by the flow
of colour quantum numbers reflecting the dynamics at short distances
(see e.g. Ref. \cite{K23}). The particle multiplicity is the result
of the convolution of the probability of primary gluon bremsstrahlung
off the original quarks with the multiplicity initiated by such a
gluon. The bulk of the QCD cascades is formed by the radiation of
quasi-collinear and/or soft gluons, with momenta
$k = (\omega; \mbf{k})$ and transverse momenta $k_{\perp}$
in the range
\begin{equation}
\mu \lesssim k_{\perp} \ll \omega \ll Q ~,
\end{equation}
where $Q$ is the hard scale of the production process.

In process (1) the standard parton showering can be generated by the
systems of quarks appearing within a short time scale,
namely the $\widehat{\t\tbar}$, $\widehat{\t\b}$ and
$\widehat{\tbar\bbar}$ antennae/dipoles \cite{K23}.
In the absence of interferences these antennae do not interact and
the particle flows are not rearranged.

The primary gluon energy, and thus the hard scale of partonic
cascades initiated by the $\widehat{\t\tbar}$ antenna, is restricted
by the kinematic condition
\begin{equation}
\omega < Q \sim \omega_{\mmax} = E_{\t} v_{\t}^2 ~,
\label{eq:kmaxt}
\end{equation}
where $v_{\t} = \sqrt{1-4\mt^2/s}$ is the top quark velocity in the
c.m. frame (in the absence of radiation) and $E_{\t} = \sqrt{s}/2$.
The virtuality scale of the parton showers corresponding to the
$\widehat{\t\b}$ and $\widehat{\tbar\bbar}$ antennae is given by the
energy $E_{\b}$ released in the top decay:
\begin{equation}
Q \sim E_{\b} \simeq \frac{\mt^2 - m_{\W}^2}{2\mt} ~.
\end{equation}

As was discussed in detail in Refs. \cite{K1,K2,K14,K15},
the energy range of primary gluons, real or virtual, generated by the
alternative quark systems of the type $\widehat{\t\bbar}$,
$\widehat{\tbar\b}$ and $\widehat{\b\bbar}$ is strongly restricted.
Not so far from the $\t\tbar$ threshold one should expect
\begin{equation}
\omega \lesssim \omega_{\mmax}^{\mrm{int}} \sim \Gt ~.
\end{equation}
Therefore the would-be parton showers initiated by such systems are
almost sterile, and can hardly lead to a sizeable restructuring of
the hadronic final state. In other words, the width of an unstable
particle acts as a kind of filter, which retains the bulk of the
radiation (with $\omega > \Gamma$) practically unaffected by the relative
orientation of the daughter colour charges.

The general analysis of soft radiation in process (1) in terms of QCD
antennae was presented in Refs. \cite{K14,K15}.
Here we focus on the emission close to the $\t\tbar$ threshold.

The primary-gluon radiation pattern can be presented as a probability
density, normalized to the lowest-order cross section:
\begin{equation}
\d N_{\g} \equiv \frac{\d \sigma_{\g}}{\sigma_0} =
\frac{\d \omega}{\omega} \, \frac{\d \Omega}{4\pi} \,
\frac{C_{\mrm{F}}\alphas}{\pi} \, {\cal I} ~,
\label{radpatt}
\end{equation}
where $\Omega$ denotes the gluon solid angle;
${\cal I}$ is obtained by integrating the absolute square of the
overall effective colour current over the virtualities of the
$\t$ and $\tbar$.

Near threshold the ${\cal O}(v_{\t}^2)$ contributions from the
$\widehat{\t\tbar}$ antenna may be neglected, and the
$\widehat{\t\b}$ and $\widehat{\tbar\bbar}$ antennae are
completely dominated by the emission off the $\b$ quarks.
In the limit of ultrarelativistic $\b$'s, the distribution
${\cal I}$ may then be presented in the form
\begin{equation}
{\cal I} = {\cal I}_{\mrm{indep}} + {\cal I}_{\mrm{dec-dec}} +
{\cal I}_{\mrm{prod-dec}} ~.
\end{equation}
Here ${\cal I}_{\mrm{indep}}$ describes the case when the $\b$ quarks
radiate independently
\begin{equation}
{\cal I}_{\mrm{indep}} =
\frac{\sin^2 \theta_1}{(1 - v_{\b} \cos \theta_1)^2} +
\frac{\sin^2 \theta_2}{(1 - v_{\bbar} \cos \theta_2)^2} ~,
\label{intindep}
\end{equation}
where $v_{\b}$ ($v_{\bbar}$) is the velocity of the $\b$ ($\bbar$)
in the lab frame, and $\theta_1$ ($\theta_2$) is the angle between
the $\b$ ($\bbar$) and the gluon.

${\cal I}_{\mrm{dec-dec}}$ corresponds to the interference between
radiation accompanying the decay of the top and of the antitop
\begin{equation}
{\cal I}_{\mrm{dec-dec}} = 2 \, \chi(\omega) \,
\frac{\cos \theta_1 \cos \theta_2 - \cos \theta_{12}}%
{(1 - v_{\b} \cos \theta_1)(1 - v_{\bbar} \cos \theta_2)} ~,
\label{intdecdec}
\end{equation}
where $\theta_{12}$ is the angle between the $\b$ and $\bbar$ and
$\chi(\omega)$ is the so-called profile function \cite{K14,K15},
which controls the radiative interferences between the different
stages of process (1). Near threshold
\begin{equation}
\chi(\omega) = \frac{\Gt^2}{\Gt^2 + \omega^2} ~.
\end{equation}

${\cal I}_{\mrm{prod-dec}}$ describes the radiative interference
between the production and decay stages:
\begin{equation}
{\cal I}_{\mrm{prod-dec}} \approx 4 \, \chi(\omega) \, v_{\t}
\left\{ \frac{\cos \Theta_1 - \cos \theta_0}%
{1 - v_{\b} \cos \theta_1} -
\frac{\cos \Theta_2 - \cos \theta_0}%
{1 - v_{\bbar} \cos \theta_2} \right\} ~.
\label{intproddec}
\end{equation}
The angles are defined as follows: $\Theta_1$ ($\Theta_2$) is the
angle between the top quark and the $\b$ ($\bbar$), and $\theta_0$
is the angle between the gluon and the top.

Let us make some comments concerning the structure of the radiation
pattern (\ref{radpatt}):
\begin{Enumerate}
\item An immediate consequence of the interference terms
${\cal I}_{\mrm{dec-dec}}$ and ${\cal I}_{\mrm{prod-dec}}$ is that
the azimuthal symmetry of the primary radiation around
the $\b$-quark direction (and thus of the particle distribution in
a $\b$ jet) is, in principle, destroyed. The resulting
skewness could be strongly dependent on the overall topology of the
final $\b\W^+ \, \bbar \W^-$ system. The interference terms can have
either sign.
\item The $\omega$-dependent coherent effects cancel after integration
over all angles. This means that the interferences do not affect the
amount of primary radiation caused by the top decays, but only
redistribute the accompanying radiation between configurations with
different relative orientation of the $\b$ jets.
\item The production--decay interference (\ref{intproddec}) vanishes
after integration over the angles between the $\widehat{\t\tbar}$ and
$\widehat{\b\bbar}$ antennae (keeping the relative angle between the
$\b$ and $\bbar$ fixed).
\item The logarithmic collinear singularities
($\sim \log E_{\b}/m_{\b}$), which dominate the integrated
distributions, appear only in the piece ${\cal I}_{\mrm{indep}}$
that describes the radiation accompanying the two separate $\t$ decays.
The interference pieces are not logarithmically enhanced.
\item Both interference terms ${\cal I}_{\mrm{dec-dec}}$ and
${\cal I}_{\mrm{prod-dec}}$ are proportional to $\chi(\omega)$. The
production--decay radiative interference is additionally suppressed
in the threshold region by the top quark velocity $v_{\t}$. (Far above
threshold the situation changes, and ${\cal I}_{\mrm{prod-dec}}$
becomes larger than ${\cal I}_{\mrm{dec-dec}}$.)
\end{Enumerate}

The profile function $\chi(\omega)$ cuts down the phase space
available for gluon emissions by the alternative quark systems and,
thus, suppresses the possibility for such systems to develop QCD
cascades. Only soft gluon emission with $\omega \lesssim \Gt$ can
lead to significant interference contributions: the radiation of
energetic gluons (either real or virtual) with $\omega \gg \Gt$
pushes the top propagators far off their resonant energy and the
interference becomes negligible.
As $\Gt \to \infty$, the top lifetime becomes very short, the
$\b$ and $\bbar$ appear almost instantaneously, and they radiate
coherently, as though produced directly. In particular, gluons from
the $\b$ and $\bbar$ interfere maximally, i.e. $\chi(\omega) = 1$.
At the other extreme, for $\Gt \to 0$, the top has a long lifetime and
the $\b$ and $\bbar$ appear in the course of the decays of colourless
top-flavoured hadrons at widely separated points in space and
time. They therefore radiate independently,
$\chi(\omega) = 0$. Thus a finite top width
suppresses the interference compared to the na\"{\i}ve expectation
of fully coherent emission. From the form of $\chi(\omega)$ we see
that the radiation pattern exhibits maximum sensitivity to $\Gt$ when
$\Gt$ is comparable to the gluon energy $\omega$. The same phenomena
appear for the interference contributions
corresponding to virtual diagrams. The infrared divergences induced
by the unobserved gluons are cancelled when both real and virtual
emissions are taken into account.

An instructive Gedanken experiment to highlight the filtering r\^ole of
$\Gamma$ can be obtained by comparing the emission of photons in
the eV to MeV range for the two processes
\begin{eqnarray}
\gamma\gamma \to \W^+\W^- & \to & \mu^+ \nu_{\mu} \,
\mu^- \overline{\nu}_{\mu}
\label{approca} ~, \\
\gamma\gamma \to \K^+\K^- & \to & \mu^+ \nu_{\mu} \,
\mu^- \overline{\nu}_{\mu}
\label{approcb} ~,
\end{eqnarray}
near threshold,
in the extreme kinematical configuration where the $\mu^+$ is
collinear with the $\mu^-$. For the first process, $\omega \ll \GW$,
and one expects hardly any radiation at all, because of the complete
screening of the two oppositely charged muons. For the second process,
$\omega \gg \Gamma_{\K}$, the parent particles have long lifetimes and
the $\mu^+$ and $\mu^-$ appear at very different times.
The photon wavelength is very small compared with the size of the
$\mu^+\mu^-$ dipole and, therefore, the $\mu^+$ and $\mu^-$
radiate photons independently, with no interference.

The bulk of the radiation caused by primary gluons with
$\omega > \Gt$ is governed by the $\widehat{\t\b}$ and
$\widehat{\tbar\bbar}$ antennae. It is thus practically unaffected
by the relative orientation of the $\b$ and $\bbar$ jets.
In particular, the $\widehat{\b\bbar}$ antenna is almost inactive.
The properties of individual $\b$ jets are understood well enough,
thanks to our experience with $\Z^0 \to \b\bbar$ at LEP 1.

Because of the suppression of energetic emission associated with
the radiative interferences, the QCD restructuring could affect only
a few particles. Therefore the magnitude $R$ of the
interconnection-induced anisotropy in the particle flow distribution
is expected to be small:
\begin{equation}
R = \frac{\Delta N_{\mrm{inter}}}{N_{\mrm{indep}}} \lesssim
\frac{1}{N_{\b}(E_{\b})} \sim \frac{1}{10} ~.
\end{equation}
Here $N_{\b}(E_{\b})$ is the multiplicity inside a $\b$ jet of energy
$E_{\b}$, corresponding to the independent emission off the $\b$
($\bbar$) quark in a $\t$ ($\tbar$) decay.

In the integral inclusive cross section for
$\ee \to \t\tbar \to \b\W^+ \, \bbar\W^-$ the rearrangement effects are
negligibly small \cite{K1}:
\begin{equation}
\frac{\Delta \sigma_{\mrm{inter}}}{\sigma} \lesssim \alphas \,
\frac{\Gt}{\mt} ~.
\end{equation}

\section{Numerical Studies}

Interconnection phenomena affect the final state of $\t\tbar$ events
in many respects, but multiplicity distributions are especially
transparent to interpret. To simplify matters, we neglect
the ${\cal I}_{\mrm{prod-dec}}$ term of eq. (\ref{intproddec});
this term is anyway small in the threshold region.

A first estimate of perturbative interconnection effects in the
primary-gluon emission rate is obtained by integrating eqs.
(\ref{intindep}) and (\ref{intdecdec}) over the solid angle
\cite{K15}. Figure~1 shows the dependence of
\begin{equation}
\int {\cal I} \, \frac{\d\Omega}{4\pi} =
\int \left( {\cal I}_{\mrm{indep}} +
{\cal I}_{\mrm{dec-dec}} \right) \frac{\d\Omega}{4\pi}
\label{intoveromega}
\end{equation}
on $\cos\theta_{\b\bbar} = \cos\theta_{12}$ for a few different
values of $\omega$. When $\omega \gg \Gt$
($\chi(\omega) \to 0$), only ${\cal I}_{\mrm{indep}}$ survives, and
so there is no dependence on $\cos\theta_{\b\bbar}$. At the opposite
limit, $\omega \ll \Gt$ ($\chi(\omega) \to 1$), the radiation
corresponds to that obtained by a $\b\bbar$ dipole.
Then radiation vanishes in the limit $\cos\theta_{\b\bbar} \to 1$,
where the $\b\bbar$ invariant mass is at threshold, while the amount
of radiation is maximal for $\cos\theta_{\b\bbar} = -1$.
Intermediate $\omega$ values give intermediate behaviours.

The non-trivial features of Fig.~1 are not experimentally testable:
only an energetic gluon ($\omega \gg 1$~GeV~$\sim \Gt$) could
be observable as an extra jet, and in this limit the
${\cal I}_{\mrm{dec-dec}}$ term is negligible. It is therefore
necessary to model the fragmentation stage and study quantities
accessible at the hadron level. This offers one advantage: the
fragmentation process has many similarities with the $\omega \to 0$
limit of the perturbative picture, and thus tends to enhance
non-trivial angular dependences. (In the limit that perturbative
radiation is neglected, the multiplicity increases logarithmically
with the $\b\bbar$ invariant mass, and thus decreases as a function
of $\cos\theta_{\b\bbar}$.)

A complication of attempting a full
description is that it is no longer enough to give the rate of
primary-gluon emission, as in eq. (\ref{radpatt}): one must also
allow for secondary branchings ($\g \to \g \g$ etc.) and specify the
colour topology and fragmentation properties of radiated partons.
It is then useful to benefit from the standard parton shower plus
fragmentation picture for $\ee \to \gammaZ \to \q\qbar$, where these
aspects are understood, at least in the sense that much of our
ignorance has been pushed into experimentally fixed parameters.

The relation between $\gammaZ \to \q\qbar$ and
$\t\tbar \to \b\W^+\bbar\W^-$ is most easily formulated in the
dipole language. The independent emission term corresponds to the
sum of two dipoles, ${\cal I}_{\mrm{indep}} \propto \widehat{\t\b} +
\widehat{\tbar\bbar}$, while the decay--decay interference one
corresponds to ${\cal I}_{\mrm{dec-dec}} \propto \chi(\omega)
(\widehat{\b\bbar} - \widehat{\t\b} - \widehat{\tbar\bbar})$.
In total, therefore,
\begin{equation}
{\cal I} \propto (1 - \chi(\omega)) \, \widehat{\t\b} +
(1 - \chi(\omega)) \, \widehat{\tbar\bbar} +
\chi(\omega) \, \widehat{\b\bbar} ~.
\label{mixthreedip}
\end{equation}
Each term here is positive definite and can be translated into a recipe
for parton shower evolution: starting from the respective normal
branching picture, each potential primary branching $\q \to \q \g$
or $\qbar \to \qbar \g$ is assigned an additional weight factor
$1 - \chi(\omega)$ or $\chi(\omega)$. This factor enters the probability
that a trial branching will be retained. For the rest, the same
evolution scheme can be used as for $\gammaZ \to \q\qbar$, including
the choice of evolution variable, $\alphas$ value, and so on.
To first approximation, this means that the $\widehat{\t\b}$ and
$\widehat{\tbar\bbar}$ dipoles radiate normally for
$\omega \gtrsim \Gt$ and have soft radiation cut off,
with the opposite for the $\widehat{\b\bbar}$ dipole.

It must be remembered, however, that soft perturbative radiation is not
that well described. This may be seen, for instance, by considering
(the 1-loop or 2-loop)
$\alphas(Q^2)$, which, at small $Q^2$ scales, is very large and rapidly
varying with $Q^2$. Therefore any moderate uncertainty in the scale
choice can have large effects on the predicted gluon emission rate.
Further, should $Q^2$ become small, the perturbative expression for
$\alphas(Q^2)$ formally blows up. Therefore it is customary in
parton shower programs to impose cuts such as
$Q^2 = \pT^2 \geq p_{\perp\mrm{min}}^2 \gg \Lambda^2$.
A typical choice is $p_{\perp\mrm{min}} \approx 0.5$~GeV.
Radiation below this scale is included in the effective fragmentation
description. There are indirect pieces of evidence for gluon
radiation down to about a 2--4 GeV mass scale \cite{TSrev},
but in the (for us important) region $\omega \leq 2 \Gt \sim 2$~GeV
the gluon emission rate given by the programs used at LEP could easily
be wrong by a factor of two without anybody ever having noticed the
difference. Therefore, while the above procedure is qualitatively
sensible, it should not be taken too literally quantitatively.

Some ambiguities also remain as to the relative time and colour ordering
of emission from the $\widehat{\t\b}$ and $\widehat{\tbar\bbar}$ dipoles,
on the one hand, and from the $\widehat{\b\bbar}$ dipole, on the other.
At first glance, it would seem natural to imagine that the former two
first radiate the more energetic gluons with $\omega \gtrsim \Gt$,
and that thereafter the energy still retained by the $\b$ and $\bbar$
defines an effective $\widehat{\b\bbar}$ dipole that radiates the
softer gluons $\omega \lesssim \Gt$. However, a closer study
of the space--time picture of gluon emission \cite{K23} reveals that
softer gluons between two energetic jets are the first to form,
followed by the radiation closer to the cores of jets. This corresponds
to a rapidity distribution of emitted gluons that builds up from the
middle outwards, in the spirit of the angular ordering phenomenon.
Therefore our standard procedure will be to begin by allowing the
$\widehat{\b\bbar}$ dipole to radiate, and thereafter use the reduced
$\b$ and $\bbar$ energies to define the respective $\widehat{\t\b}$
and $\widehat{\tbar\bbar}$ dipole radiation. This has the added
advantage, using the standard parton shower routines \cite{TS2},
of giving a sensible colour ordering between emitted partons:
$\b$---(gluons from the $\widehat{\t\b}$ dipole)---(gluons from the
$\widehat{\b\bbar}$ dipole)---(gluons from the $\widehat{\tbar\bbar}$
dipole)---$\bbar$. With the opposite ordering, some gluons from the
$\widehat{\b\bbar}$ dipole will sit next to the $\b$ and the others next
to the $\bbar$, which is less physical. However, fluctuations will
obviously occur, which break any assumed strict ordering and lead to an
interleaving of emission from the three dipoles. Both alternatives
above will therefore be studied to assess the amount of uncertainty
that could be induced by our imperfect understanding of colour ordering.

There is one important difference between the $\widehat{\t\b}$ dipole
and the $\widehat{\b\bbar}$ one, namely how the recoil of gluon emission
is distributed. The $\widehat{\t\b}$ dipole consists of a back-to-back
$\b$ and $\W^+$ in the rest frame of the $\t$. The energy of the $\W$
is given by $E_{\W} = (m_{\t}^2 + m_{\W}^2 - m_{\b^*}^2) / 2 m_{\t}$,
where $m_{\b^*}$ gives the invariant mass of the $\b$ and all the
gluons emitted. That is, the more the radiation, the more of the
energy goes into the hadronic final state and the less is left for the
leptonic decay products of the $\W$. Gluon emission from the
$\widehat{\b\bbar}$ dipole, on the other hand, does not change the
total hadronic energy, but only redistributes energy between the $\b$,
the $\bbar$ and the emitted gluons.

In total, we have chosen to compare the multiplicity distributions of
six different scenarios:
\begin{Enumerate}
\item No QCD radiation at all. A simple string is stretched between
the $\b$ and $\bbar$ and thus fragmentation introduces a dependence
of particle production on $\cos\theta_{\b\bbar}$.
\item QCD radiation from the $\widehat{\b\bbar}$ dipole only,
with full strength (i.e. no $\chi(\omega)$ dampening factor).
Both the perturbative and non-perturbative stages now depend
on $\cos\theta_{\b\bbar}$.
\item Formation of two independent top-flavoured hadrons, which
each radiate (as ordinary $\widehat{\t\b}$ and $\widehat{\tbar\bbar}$
dipoles) in their decays. There is then no dependence on
$\cos\theta_{\b\bbar}$.
\item QCD radiation from the $\widehat{\t\b}$ and
$\widehat{\tbar\bbar}$ dipoles only, with full strength
(i.e. no $1 - \chi(\omega)$ dampening factor). As in
scenario 1, fragmentation introduces a dependence on
$\cos\theta_{\b\bbar}$.
\item QCD radiation from all three dipoles, according to eq.
(\ref{mixthreedip}), with the $\widehat{\b\bbar}$ dipole radiation
considered last.
\item QCD radiation from all three dipoles, with the
$\widehat{\b\bbar}$ dipole radiation considered first.
This is our preferred scenario.
\end{Enumerate}
The alternatives are given roughly in order of increasing realism,
with the first three mainly intended to illustrate various extreme
behaviours.

In all the above cases, the top quarks are assumed to decay
isotropically in their respective rest frame, i.e. we do not
attempt to include spin correlations between $\t$ and $\tbar$.
Close to threshold this gives an essentially flat distribution
in $\cos\theta_{\mrm{parton}}$, defined as the angle between
the `original' $\b$ and $\bbar$ directions before QCD
radiation effects are considered.
(For the points we are trying to make, the input angular
distribution is not of interest, but only the variation of event
properties with angle.) Breit-Wigner distributions are
included for the top and $\W$ masses.

The dependence of the average charged multiplicity on
$\cos\theta_{\mrm{parton}}$ is shown in Fig.~2 for the six scenarios
defined above. The top mass is 160 GeV and the c.m. energy 330~GeV,
i.e. 10 GeV above threshold.
In scenario 3 top hadrons are allowed to form,
and the multiplicity in the subsequent decays is therefore independent
of decay angles. When the top mass was thought to be smaller, i.e. up
until a few years ago, this would have been the natural behaviour.
(For illustration, we show results for $m_{\T} = m_{\t} + m_{\q}$, with
$m_{\u} = m_{\d} \approx 0.3$~GeV; a larger light-quark mass would give
a larger multiplicity. This uncertainty is the price to be paid for
introducing intermediate states in the process, and is related to the
issue of how to define the heavy-flavour mass in a bound state.)
In all the other scenarios, a perturbative
and/or non-perturbative interconnection is introduced between the
$\b$ and $\bbar$ jets, and then the total multiplicity is smaller, the
smaller the angle between the two. The dependence is especially
spectacular in scenario 2, where the $\widehat{\b\bbar}$ dipole
is allowed also to radiate energetic gluons. If the energetic gluon
radiation instead comes from the $\widehat{\t\b}$ and
$\widehat{\tbar\bbar}$ dipoles, the effects of soft gluons and
fragmentation give a rather more modest variation with
$\cos\theta_{\mrm{parton}}$, although still easily visible.

The $\theta_{\mrm{parton}}$ angle is a desktop product, not an
experimentally accessible quantity. One realistic alternative
would be to use microvertex detectors to tag the $\B$ hadron
directions. The $\B\Bbar$ relative angle would then differ from
$\theta_{\mrm{parton}}$ by a number of effects, such as the recoil
from gluon radiation. A simpler alternative is to reconstruct exactly
two clusters per event, leaving aside the (leptonic) decay
products of the two $\W$'s, and consider the angle between the two
clusters. The resulting $\cos\theta_{\mrm{cluster}}$ distribution is
shown in Fig.~3. Scenarios 1 and 2 here display one of the well-known
features of the string/drag phenomenon, namely that two
colour-connected partons together produce particles in the region
in between the two, so that jet axes seem to sit closer in angle
than they really should do \cite{TSJADE}. The other scenarios show the
opposite trend, where jets seemingly are `repelled' from each other.
However, what is going on is a bit more complex than that.
In these scenarios energetic gluons can be emitted anywhere in angle,
not just in between the $\b$ and $\bbar$. If three (or more) separated
jets are present, the clustering procedure will join the two closest
into one (even when that means that both the $\B$ and $\Bbar$
hadrons come to belong to the same cluster), and therefore deplete the
low-$\theta_{\mrm{cluster}}$ region.

The dependence of the average charged multiplicity on
$\cos\theta_{\mrm{cluster}}$ is shown in Fig.~4. The differences
between the models are qualitatively the same as in Fig.~2,
although the variation with relative angles is now more pronounced
in scenarios 3--6; in particular, remember that scenario 3 in flat
in Fig.~2. The reason is clear: the `repulsion' effect mentioned above
is predominantly active for events that contain energetic gluon
emission and therefore above-average multiplicities. The region
of small $\theta_{\mrm{parton}}$ then predominantly contains
two-jet, low-multiplicity events. If one so wished, cuts on jet
structure and knowledge of $\B$ microvertices could
be used to obtain results closer to Fig.~2, while Fig.~4 represents
the separation between models that is visible already with a minimum
of effort.

In order to test the observability of the features of Fig.~1,
the charged particle multiplicity may be split into momentum bins.
Low-momentum particles, like low-momentum gluons, should feel
the interconnection effects more. In this region the multiplicity
dependence on $\theta_{\mrm{cluster}}$ (or $\theta_{\mrm{parton}}$)
should therefore be more pronounced. This is indeed the case,
see Fig.~5. The dependence on $\cos\theta_{\mrm{cluster}}$
in scenario 3 is entirely due to the `repulsion' effect noted above;
for $\cos\theta_{\mrm{parton}}$ this distribution is essentially flat.
(Actually, it is even slightly going up with
$\cos\theta_{\mrm{parton}}$; this is a kinematical reflection of the
top hadrons not being quite at rest, and is compensated by
higher-momentum particles.) At larger momenta the distributions
progressively flatten out, i.e. energetic hadrons better follow
the directions of the original partons. There is also a
momentum-conservation effect, that events (and scenarios) with fewer
low-momentum particles have more high-momentum ones.

The behaviour of scenarios 4--6 is sufficiently well separated from
the other three to allow easy experimental tests. Since the
perturbative soft-gluon emission and the non-perturbative
fragmentation are seen to be working in the same direction,
the differences between scenarios 4, 5 and 6 are much smaller.
The sign of the differences is understood, e.g. the admixture of the
$\widehat{\b\bbar}$ dipole leads to a variation of the multiplicity
with $\cos\theta$ larger in scenario 5 than in scenario 4,
but the absolute magnitude can be questioned, as we have noted before.
It is here premature to promise a well-defined
experimental separation, but possibly this could be achieved
in the end. For instance, so far we only considered the inclusive
multiplicity, not its distribution relative to the jet directions,
which can provide further handles.

The optimal determination of the top mass is a complex task, and
we do not attempt a complete analysis here. However, a few general
comments can be made. We have considered two approaches to the
$\mt$ determination: either to use the lepton spectrum to reconstruct
a $\W$ momentum distribution, or to form $\W\b$ invariant masses
by a subdivision of the hadronic final state. For the former method,
scenarios 1 and 2 both give
$\langle p_{\W} \rangle \pm \sigma(p_{\W}) = 66.0 \pm 15.3$~GeV,
scenario 3 gives $63.0 \pm 15.4$~GeV, and scenarios 4--6
give $63.4 \pm 15.8$~GeV. That is, the $\widehat{\t\b}$ and
$\widehat{\tbar\bbar}$ dipole radiation reduces the average $\W$
momentum significantly, and top hadron formation reduces it further.
By comparison, an increase of the top mass by 1~GeV, for a
fixed c.m. energy, reduces the average $\W$ momentum by 0.3~GeV
and narrows the width by 1.1~GeV.
For the second method, we have made the (unrealistic) assumption
that the two $\W$ momenta are perfectly known, so that the key task
is to split the rest of the hadronic system into a $\b$ cluster and
a $\bbar$ one. Each $\b$ is then paired with a $\W$ in the combination
that gives the most `reasonable' set of two masses. The average
reconstructed top momentum (the two need not be the same, since
some neutrinos are occasionally lost by semileptonic $\b$ decays)
can be used to
reconstruct an average top mass of the event. This average mass comes
out to be 159.3~GeV in scenario~1, 159.1~GeV in scenario~2,
158.8~GeV in scenario~3, and 158.5~GeV in scenarios~4--6.
This is to be compared with the true 159.2~GeV, lower than the nominal
160~GeV because of the balance between Breit-Wigners and phase space.
The mass shift in the latter scenarios is likely to be related to the
`repulsion' effect noted above. The conclusion is that the usage of an
unrealistic scenario could give an error on the top mass determination
by half a GeV. However, the separation between scenarios 4, 5 and 6
comes out to be of the order of 30~MeV, and should therefore
constitute a negligible uncertainty.

Since the amount of soft-gluon emission is significantly reduced by the
shower cut-off, we have repeated the whole procedure for a toy model
with a pure parton-level simulation of gluon emission, in the spirit of
eq.~(\ref{radpatt}), with a fixed $\alphas = 0.25$ and a lower cut-off
$\omega_{\mrm{min}} = 0.01$~GeV. This allows us to assess the potential
importance of neglected low-momentum gluons, but should otherwise not
be taken seriously. This simpler model qualitatively
agrees very well with all the relevant features discussed above,
both for multiplicity flows and for top-mass determinations.
In particular, the reconstructed top mass differs by an order
of 20~MeV between scenarios 4, 5 and 6. The conclusion seems to be that
the task of subdividing the $\b\bbar$ hadronic system into two clusters
is a crude one, not particularly sensitive to the finer details of the
event such as the ones shown in Figs.~2--5.

As a final comment, we note that one nice signal for interconnection
effects would be the presence of an $\Upsilon$ in the final state,
when the $\b$ and $\bbar$ from the $\t$ and $\tbar$ decays have such
a small invariant mass that the hadronic system collapses to an
exclusive final state. Unfortunately, the probability for this to
occur comes out very small, below $10^{-3}$ for scenarios 1 and 2
and below $10^{-4}$ for scenarios 4--6. It is therefore not clear
whether this aspect can be studied experimentally.

\section{Conclusion}

In $\ee \to \t\tbar \to \b\W^+ \, \bbar\W^-$ events the QCD
interconnection phenomena can rearrange the particle flows and
for example produce particles that cannot be uniquely assigned to
either $\t$ or $\tbar$. We have shown that, on the
perturbative level, this rearrangement is suppressed.
Further, we have estimated the non-perturbative effects.
It is not our aim here to present a full-scale treatment of
the final-state structure in $\ee \to \t\tbar$ in the whole
energy range of a future linear collider. It would be rather
premature to perform a too detailed analysis of the events
that may be produced ten years from now. Instead we concentrated
here on the relatively simple case of
$\t\tbar \to \b\W^+ \, \bbar\W^-$ production a few GeV above
the top threshold.

On the experimental side, the main conclusions of our study are:
\begin{Itemize}
\item The interconnection phenomena should be readily visible in the
variation of the average multiplicity as a function of the relative
angle between the $\b$ and $\bbar$ clusters.
\item A more detailed test is obtained by splitting the particle
content in momentum bins. The high-momentum particles are mainly
associated with the $\widehat{\t\b}$ and $\widehat{\tbar\bbar}$
dipoles and therefore follow the $\b$ and $\bbar$ directions,
while the low-momentum ones are sensitive to the assumed influence
of the $\widehat{\b\bbar}$ dipole.
\item An incorrect simulation of energetic gluon emission could lead
to significant errors in top mass determinations. However, the
effects of the $\widehat{\b\bbar}$ dipole
(when dampened by $\chi(\omega)$, i.e. when only affecting
low-momentum gluon emission) on the reconstructed $m_{\t}$ seem
to be negligible. Therefore, leaving aside a host of uncertainties
not studied in this paper, we have demonstrated that a correct
description of the event shapes in top decay, combined with
sensible reconstruction algorithms, should give errors on the
top mass that are significantly less than 100~MeV.
\end{Itemize}

The possibility of interference rearrangement effects in the
$\t\tbar$ production is surely not restricted to the events
studied here. One could discuss also hadronic $\W$ decays and/or the
interferences with beam jets in $\p\p / \p\pbar \to \t\tbar$
events. The problem with these processes is that there are too
many other uncertainties that make systematic studies look
extremely difficult. At the moment, the main uncertainties come
from the modelling of the non-perturbative fragmentation, as
for the study of this paper. However, the main conclusion that
the QCD interconnection does not induce any sizeable
restructuring of the final particle flows should remain valid
for all of these cases.

\subsection*{Acknowledgements}

This work was supported in part by the UK Science and Engineering
Research Council. The authors are grateful to Yu.L. Dokshitzer and
W.J. Stirling for useful discussions.

\clearpage

\section*{Figure captions}

\begin{list}{}{\setlength{\leftmargin}{2cm}
  \setlength{\labelwidth}{1.3cm}\setlength{\labelsep}{0.7cm}
  \setlength{\rightmargin}{0cm}}

\item[Fig.~1]
Integrated gluon emission rate, eq. (\ref{intoveromega}), as a
function of $\cos\theta_{\b\bbar}$. Full curve is for $\omega = 0$,
dashed for $\omega = \Gt/2$, dotted for $\omega = \Gt$,
dash-dotted for $\omega = 2\Gt$, and the full straight line for
$\omega \to \infty$.  Note that the prefactor
$C_{\mrm{F}}\alphas/\pi\omega$ of eq.~(\ref{radpatt}) has been
omitted so as to emphasize the dependence on $\cos\theta_{\b\bbar}$
rather than the variation with $\omega$.

\item[Fig.~2]
Average charged multiplicity as a function of
$\cos\theta_{\mrm{parton}}$. Scenario 1 dotted, scenario 2 dash-dotted,
scenario 3 filled triangles, scenario 4 dashed,
scenario 5 open squares, and scenario 6 (the preferred one) full.
Note that the zero has been suppressed on the $y$ scale.

\item[Fig.~3]
Distribution of $\cos\theta_{\mrm{cluster}}$. Notation as in Fig.~2.

\item[Fig.~4]
Average charged multiplicity as a function of
$\cos\theta_{\mrm{cluster}}$. Notation as in Fig.~2.
Note that the zero has been suppressed on the $y$ scale.

\item[Fig.~5]
Average multiplicity of charged particle with momenta
$|\mbf{p}| < 1$~GeV as a function of
$\cos\theta_{\mrm{cluster}}$. Notation as in Fig.~2.

\end{list}

\end{document}